\documentclass[12pt]{iopart}
\usepackage{iopams}  
\usepackage{graphicx,color}
\usepackage{draftcopy}
\usepackage[squaren]{SIunits}
\usepackage{color}      
\usepackage{hyperref}   

\definecolor{mygreen}{rgb}{0,0.5,0} 
\definecolor{myblue}{rgb}{0,0,0.75} 
\definecolor{mymagenta}{cmyk}{0,1,0,0.12} 

\newcommand{\gtext}[1]{{\color{mygreen}#1}}

\newcommand{\ctext}[1]{{\color{cyan}#1}}

\renewcommand{\gtext}[1]{{#1}}

\renewcommand{\ctext}[1]{{#1}}

\newcommand{\bV}{{\bf{V}}}
\newcommand{\Vhat}{\hat{\bf{V}}}
\newcommand{\Jhat}{\hat{\bf{J}}}
\newcommand{\Shat}{\hat{\bf{S}}}
\newcommand{\Fhat}{\hat{\bf{F}}}
\newcommand{\npulse}{N_{\rm pulses}}
\newcommand{\supi}{^{(m)}}

\newcommand{\FPlane}{{F_{||}}}

\newcommand{\var}{{\rm var}}
\newcommand{\xitwomin}{\xi^{2}_{\rm min}}
\newcommand{\supin}{^{({\rm in})}}
\newcommand{\supout}{^{({\rm out})}}
\newcommand{\cov}{{\rm cov}}

\begin{document}

\title{Planar  squeezing {by} quantum non-demolition measurement in cold atomic ensembles}

\author{Graciana Puentes$^{1}$, Giorgio Colangelo$^{1}$, Robert J. Sewell$^{1}$, and Morgan W. Mitchell$^{1,2}$}

\address{$^{1}$ ICFO - Institut de Ciencies Fotoniques, Mediterranean Technology Park, 08860 Castelldefels, Barcelona, Spain\\
\noindent $^{2}$ ICREA - Institucio Catalana de Recerca i Estudis Avancats, 08015 Barcelona, Spain}
%
%

\ead{morgan.mitchell@icfo.es}

\newcommand{\pqs}{planar squeezing }

\begin{abstract}
\noindent Planar  squeezed states, i.e. quantum states which are squeezed in two orthogonal spin components, have recently attracted attention due to their applications in atomic interferometry and quantum information [Q. Y. He et al, New J. Phys. {\bf 14}, 093012 (2012)].~While canonical variables such as quadratures of the radiation field can be squeezed in at most one component, simultaneous squeezing in two orthogonal spin components can be achieved due to the angular momentum commutation relations.~We present a novel scheme for \pqs via quantum non-demolition (QND) measurements in spin-1 systems. The QND measurement is achieved via near-resonant paramagnetic Faraday rotation probing, and the \pqs is obtained by sequential QND measurement of two orthogonal spin components.~We compute the achievable squeezing for a variety of optical depths, initial conditions, and probing strategies.  The planar squeezed states generated in this way contain entanglement detectable by spin-squeezing inequalities and {give an advantage relative to non-squeezed states for any precession phase angle, a benefit for single-shot and high-bandwidth magnetometry.}
\end{abstract}
%
\maketitle

\section{Introduction}

Quantum squeezing, the reduction of uncertainty in quantum observables below ``classical'' or standard quantum limits \cite{CavesPRD1981,Wineland1992,KitagawaPRA1993}, is an area of active research with applications in quantum information processing \cite{BraunsteinBOOK2010,OurjoumtsevS2006,DubostPRL2012,ChristensenNJP2013,BeduiniARX2013a}, quantum communications \cite{VahlbruchNJP2007,AppelPRL2008,UsenkoNJP2011}, and quantum metrology \cite{GiovannettiS2004,WolfgrammPRL2010,SewellPRL2012}.  Here we consider a practical, measurement-based strategy to produce a new kind of squeezing in atomic spin ensembles.  By way of introduction, we note that a spin ${\bf F}$ obeys the spin uncertainty relations
\begin{equation}
\label{eq:SpinUncRelation}
\Delta F_{z} \Delta F_{x} \geq \frac{1}{2} {|\langle F_y \rangle|}
\end{equation}
and permutations (throughout we take $\hbar = 1$).  These are distinctly different from the Heisenberg relation $\Delta X \Delta P~\geq~\frac{1}{2} $ that governs canonical variables $X,P$ ~\cite{LoudonJMO1987}.  For canonical variables,  the constant right hand side (RHS) of the Heisenberg relation implies that reduction of the variance in one quadrature inevitably  increases the variance of the other.  In contrast, in spin systems the RHS of the uncertainty relation may vanish, e.g. if $\langle F_{y} \rangle =0$, with the consequence that two spin components, e.g. $F_z$ and $F_x$, may be {\it simultaneously} squeezed, with the uncertainty being absorbed by the third component.  An analogous situation involving the Stokes parameters of light, which themselves obey angular momentum commutation relations, has also been studied~\cite{KorolkovaPRA2002,SchnabelPRA2003,Predojevic2008}.  For atomic spin ensembles, the suggestion of ``intelligent'' spin states~\cite{AragoneJPA1974} predates even the concept of optical squeezing, and {\em planar quantum squeezing} (PQS) has recently been analyzed and proposed for quantum metrological applications~\cite{HePRA2011,HeNJP2012}.

\begin{figure}[ht!]
\centering
\includegraphics[width=0.5\textwidth]{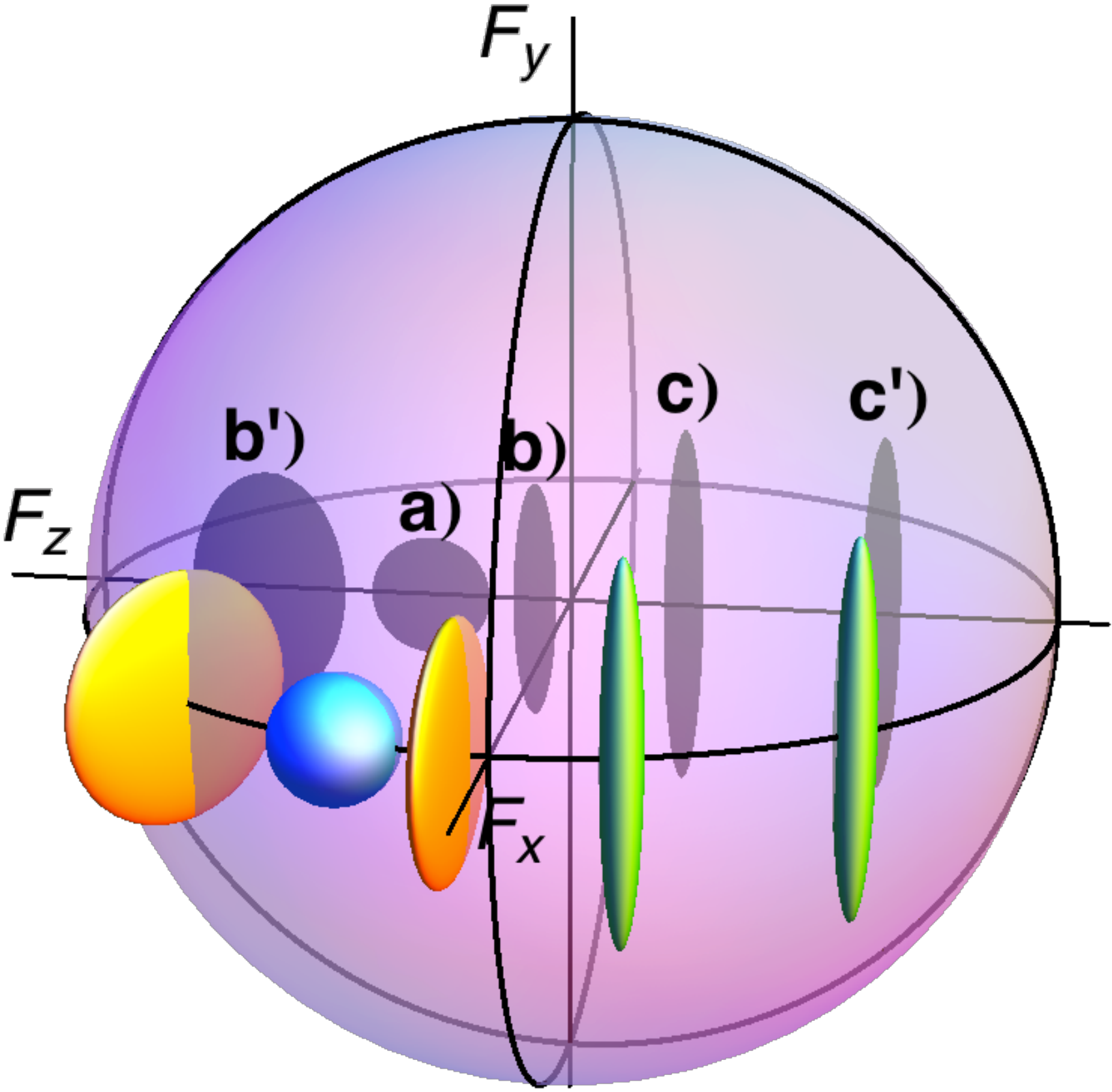}
\caption{Illustration of spin variances for: a) an unsqueezed state with SQL variances $\Delta^2 F_{i}$ for all components i $\in \{x, y, z\}$, b), b$'$) a state with squeezed azimuthal component.
For measurements of $F_{z}$, this squeezing is advantageous only for certain precession angles, as in b), but disadvantageous for others, as in b$'$).~c), c$'$) a planar squeezed state with both variances $\Delta^2 F_{x}$ and $\Delta^2 F_{z}$ reduced below the SQL is advantageous at all angles.
\label{fig:3DVariances}}
\end{figure}

As described in~\cite{HePRA2011,HeNJP2012},~PQS has the prospect of improving the precision of atomic interferometers at arbitrary phase angles, as it allows both orthogonal variances to be below the shot-noise level \cite{BerryPR2009, HigginsJP2009}.~This is opposed to squeezing on a single direction, which is only beneficial to refine the estimate of a phase angle already known with some precision (see Fig.~1).~High-bandwidth atomic magnetometry~\cite{ShahPRL2010,VasilakisPRL2011}, in which the precession angle may not be predictable in advance, would be a natural application.  At the same time, there is interest in the entanglement properties of PQS states, which can be detected by spin-squeezing inequalities~\cite{HePRA2011,HeNJP2012,VitaglianoPRL2011}.

He {\it et al.} proposed an implementation of PQS using a two-well Bose condensate with tunable and attractive interactions~\cite{HePRA2011}.~Here,~we propose an alternative strategy in which the PQS state is prepared by sensitive \cite{KoschorreckPRL2010a} non-destructive \cite{KoschorreckPRL2010b} spin measurements.~Sufficiently good measurements satisfy quantum non-demolition (QND) criteria \cite{SewellARX2013} and produce (conditional) squeezing \cite{SewellPRL2012} of the measured spin variable.~Stroboscopic probing a polarized ensemble of cold $^{87}$Rb atoms, precessing in a constant external magnetic field, allows sequential measurement of two orthogonal components.~As we show, this can squeeze two spin components simultaneously.~We work with the collective total angular momentum variable ${\bf F}$ for spin-1 systems.~The collective atomic spin is measured using paramagnetic Faraday rotation with an off-resonant probe.~The ensemble of spins, polarized along a fixed direction by optical pumping, interacts with an optical pulse of duration $\tau$ and polarization described by the Stokes vector $\textbf{S}=(S_{x},S_{y},S_{z})$, through an effective Hamiltonian.\\

The article is structured as follows: In Section 2, we describe the formalism for light-atom interactions in polarized atomic ensembles and define spin-squeezing parameters. In Section 3, we report numerical studies of PQS by QND measurement. In Section 4, we discuss entanglement detection using spin-squeezing inequalities for the generated PQS states.  In Section 5  we provide an example for the applications of these ideas in high bandwidth (single shot) atomic magnetometry. 

\section{Atom-light Interaction}

Our modeling is based on covariance matrix techniques introduced by Madsen and M{\o}lmer~\cite{MadsenPR2004,EchanizJO2005,MolmerPRA2004}, extended to include inhomogeneities by Koschorreck {\it et al.}~\cite{KoschorreckJPAMOP2009}, to three spin components by Toth {\it et al.} \cite{TothJP2010} and to spin-1 atoms by Colangelo {\it et al.}~\cite{ColangeloARX2013}.~In brief, we consider a magnetic field ${\bf B}$  oriented along the $y$ axis, in interaction with a spin-1 ensemble, e.g. $^{87}$Rb in the $F=1$ ground state, and probed using Faraday rotation probing by near-resonant light pulses.  The ensemble is described by collective spin observables, for example  \ctext{$\hat{F}_x \equiv \sum_i \hat{f}_x^{(i)}$}, where $\hat{f}_x^{(i)}$ indicates the $x$-component of the spin of the $i$'th atom.~Here $F_{i}$ correspond to the macroscopic magnetization components, which precess about the external magnetic field $\textbf{B}$ at the Larmor frequency.~Both  spin orientation variables $\hat{F}_x, \hat{F}_y,\hat{F}_z$, and also spin alignment variables $\hat{J}_x,\hat{J}_y,\hat{J}_k,\hat{J}_l,\hat{J}_m$ defined using  $\hat{j}_x \equiv \hat{f}_x^2 - \hat{f}_y^2$, $\hat{j}_y \equiv \hat{f}_x \hat{f}_y + \hat{f}_y \hat{f}_x$, $\hat{j}_k \equiv \hat{j}_k =\hat{f}_x\hat{f}_z+\hat{f}_z \hat{f}_x \quad \hat{j}_l =\hat{f}_y \hat{f}_z+\hat{f}_z \hat{f}_y \quad  \hat{j}_m=\frac{1}{\sqrt{3}}(2\hat{f}_z^2-\hat{f}_x^2-\hat{f}_y^2)$, respectively, are required to describe the spin-1 system.~The optical pulses are described in terms of Stokes operator components $\hat{S}_x,\hat{S}_y, \hat{S}_z$, defined as:
\begin{equation}
\label{eq:StokesDef}
\hat{S}_x\equiv\frac{1}{2}\hat{\mathbf{a}}^{\dag}\mathbf{\sigma}_x\hat{\mathbf{a}},
\quad \hat{S}_y\equiv\frac{1}{2}\hat{\mathbf{a}}^{\dag}\mathbf{\sigma}_y\hat{\mathbf{a}},
\quad \hat{S}_z\equiv\frac{1}{2}\hat{\mathbf{a}}^{\dag}\mathbf{\sigma}_z\hat{\mathbf{a}},
\end{equation}
where $\hat{\mathbf{a}}\equiv(\hat a_+,\hat a_-)^T$ and $\hat a_+, \hat a_-$ are the annihilation operators for the left and right circular polarization and $\mathbf{\sigma}_x$, $\mathbf{\sigma}_y$, $\mathbf{\sigma}_z$ the Pauli matrices.~The $\hat{S}_x$, $\hat{S}_y$ and $\hat{S}_z$ Stokes operators represent polarized light in the horizontal or vertical direction, polarized light in the $\pm 45^o$ direction, and right-hand or left-hand circularly polarized light, respectively.~As mentioned, they obey the same commutation relations as angular momentum operators.

The full computational machinery is described in detail in Ref.~\cite{ColangeloARX2013}.  
In brief, we use a phase-space vector to describe the state of the whole system
\begin{eqnarray}
\Vhat = {\bf B} \oplus {\Fhat} \oplus {\Jhat} \oplus \bigoplus_{i=1}^{\npulse} {\Shat}\supi 
\label{eq:stateVector}
\end{eqnarray}
where $\oplus$ indicates the direct sum, the superscript $\supi$ indicates the $m$'th optical pulse, and ${\bf B}$ is the magnetic field vector at the location of the ensemble.  ${\bf B}$ is here a classical field, whereas the other components of $\Vhat$ are operators.~We work within the gaussian approximation, i.e., we assume that $\Vhat$ is fully characterized by its average $\langle{\Vhat}\rangle$ and by its covariance matrix  $\Gamma_\bV$:
\begin{eqnarray}
\label{GammaV}
 \Gamma_\bV\equiv\frac{1}{2}\langle\Vhat\wedge\Vhat+(\Vhat\wedge\Vhat)^T\rangle-\langle\Vhat\rangle\wedge\langle\Vhat\rangle
\end{eqnarray}
where $\wedge$ indicates the outer product.~The phase space vector $\Vhat$ evolves under the effect of the hamiltonian
\begin{equation}
\label{eq:Heff}
H_{\mathrm{eff}} = -g_{F} \mu_{B} {\bf B}{\cdot} {\hat{\bf F}}+\frac{1}{\tau}[G_{1}\hat{S}_{z} \hat{F}_{z} + G_{2}(\hat{S}_{x}\hat{J}_{x}+\hat{S}_{y}\hat{J}_{y})],
\end{equation}
where $\mu_{B}$ is the Bohr magneton and $g_{F}$ is the gyromagnetic factor,  $\tau$ is the pulse duration, and $G_{1,2}$ are vectorial and tensorial coupling coefficients, respectively.~Also included in the evolution are the decohering effects of incoherent scattering of probe photons and inhomogeneous magnetic fields~\cite{KoschorreckJPAMOP2009}.

\begin{figure}[ht!]
\centering
\includegraphics[width=0.8\textwidth]{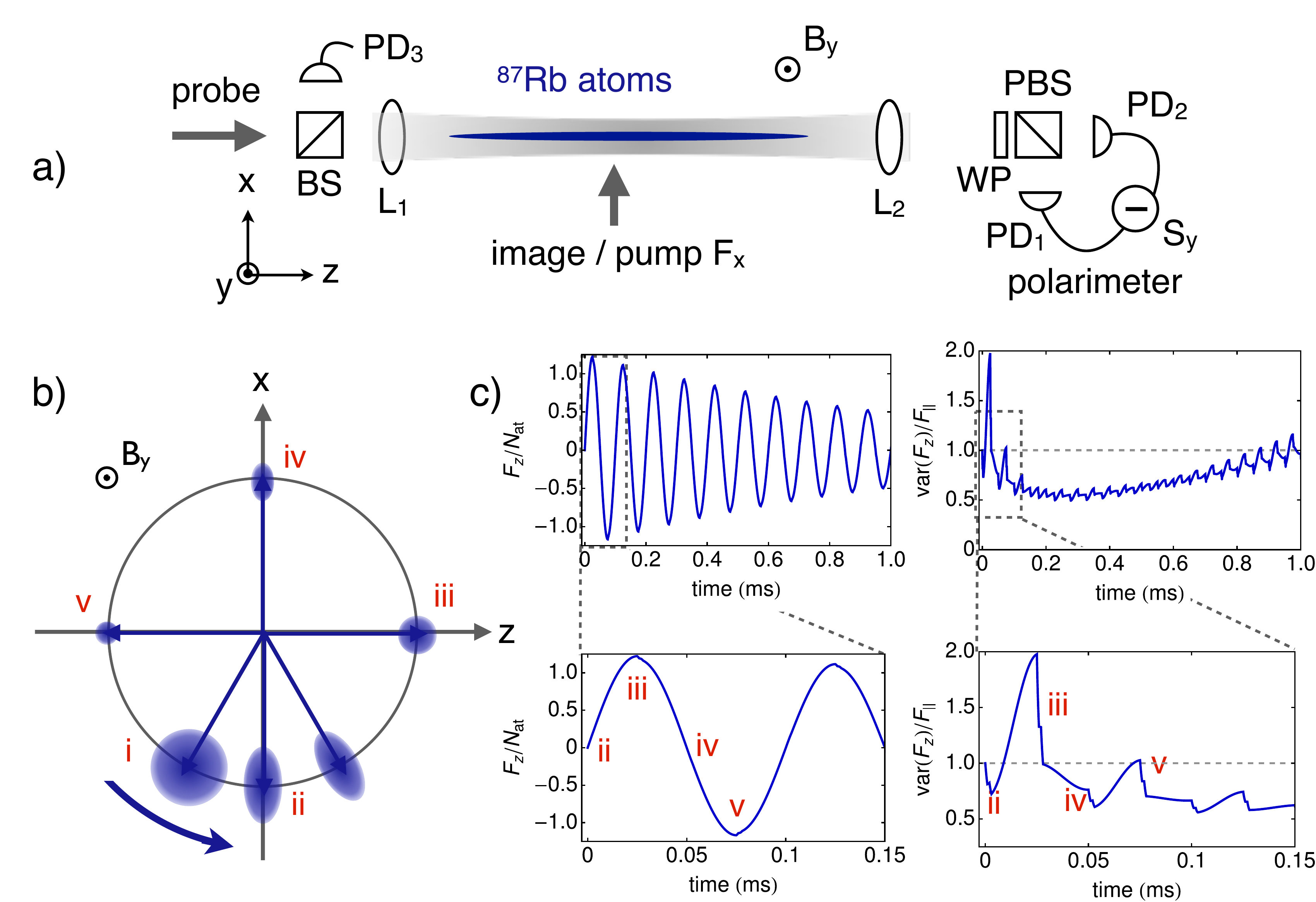}
\caption{
(a) Experimental setup.  PD: photodiode; L: lens; WP: wave plate; BS: beam splitter; PBS: polarizing beam splitter.
(b) \& (c) \pqs is achieved by measuring the Faraday rotation of the Stokes components in the off-resonant probe light at four different moments during the Larmor precession cycle ~\cite{NapolitanoN2011}.
Each measurement consists of a pair of h-- and v--polarized optical pulses and has a total duration of 3~$\mu$s, and produces a Faraday rotation signal proportional to the $z$--component of the collective atomic spin in the laboratory frame of reference at that phase of the 100~$\mu$s long Larmor precession cycle.
In (b) we illustrate the measurement cycle: (i) a PCSS is prepared and rotates in the $x$--$z$ plane due to the applied magnetic field $B_y$.
The first measurement (ii) squeezes the $F_z$ component of the collective spin.
The second measurement (iii) squeezes the $F_x$ component of the collective spin, which is now aligned along the $z$--axis of the laboratory frame.
The two measurements are repeated at moments (iv) and (v), further squeezing the $F_z$ and $F_x$ components of the collective atomic spin.
The evolution of the average collective spin $F_z$ and variance var($F_z$) in the laboratory frame of reference during this measurement cycle is illustrated in (c).
The top left panel shows the precession of the average $F_z$ over multiple Larmor precession periods.
The collective spin coherence decays due to scattering of photons during the measurements.
The moment that the measurements (ii)--(v) are made is indicated in the bottom left panel, which shows a magnification of the first Larmor precession cycle.
The evolution of the variance var($F_z$) of the collective spin is shown in the top right panel.
The sharp jumps are due to the measurement pulses, which squeeze the measured $F_z$ component (in the laboratory frame).
In between the pulses noise is rotated into the $F_z$ component of the collective spin by the magnetic field.
A magnification of the first Larmor precession cycle is shown in the bottom right panel.
Pulses (ii) \& (iv) squeeze the $F_z$ component (in the atomic frame) and pulses (iii) \& (v) squeeze the $F_x$ component.
\label{fig:ExperimentalScheme}}
\end{figure}

The strategy to generate planar squeezing is illustrated in Figure~\ref{fig:ExperimentalScheme}, and can be described as follows:~An initial state is prepared and made to precess in the $x$--$z$ plane by a magnetic field oriented along $y$ (the ${\bf B}{\cdot} {\hat{\bf F}}$ term in $H_{\mathrm{eff}}$).~Pulses of light, short on the time-scale of the precession, and with initial polarization along $S_x$, are sent through the ensemble at quarter-cycle intervals.~Through the $\hat{S}_{z} \hat{F}_{z}$ term, the light-atom interaction rotates the polarization from $S_x$ toward $S_y$ by an angle proportional to $\hat{F}_z$.~A measurement of $S_y$ then gives an indirect measurement of $F_z$, reducing its uncertainty.~If this measurement is sufficiently precise it leaves $F_z$ squeezed.~A quarter cycle later, the state has precessed so that the $F_x$ component can be measured, and possibly squeezed \cite{SewellARX2013}.~The  terms proportional to $G_2$ couple the orientation ($\bf F$) components to the alignment ($\bf J$) components, and for this application are simply an inconvenience.~We suppress the effect of the $G_2$ term by dynamical decoupling, as described in Reference~\cite{KoschorreckPRL2010b}. 

While a given spin component can be squeezed by non-demolition measurement of that component, it is perhaps not obvious that planar squeezing, which requires simultaneous squeezing of non-commuting observables, should be producible by measurement.~There is, after all, no projective measurement that indicates {\em both} $F_x$ and $F_z$.  Nevertheless, if $\langle F_y \rangle \approx 0$, the uncertainty relation of Eq.~(\ref{eq:SpinUncRelation}) does not impose a quantum back-action on $F_x$ when $F_z$ is measured, nor vice-versa.  This suggests that sequentially measuring $F_x$ and $F_z$ should squeeze both spin components and thus give planar squeezing. 

The operational definition we adopt for \pqs follows the approach by He {\em et al}.~\cite{HePRA2011,HeNJP2012}, and can be summarized as follows. First, from Eq~(\ref{eq:SpinUncRelation}) and permutations, we take  $\Delta^2 F_x = \Delta^2 F_z = \FPlane/2$ as the SQL, where $\FPlane \equiv \sqrt{F_x^2 + F_z^2}$, so that $\FPlane$ is the magnitude of the in-plane spin components.  Considering then the noise in the plane, we define the planar variance $\Delta^2 \FPlane \equiv  \Delta^2 F_{x} + \Delta^2 F_{z}$, with SQL $\Delta^2 \FPlane = \FPlane$, and  the {\em planar squeezing parameter}
\begin{equation}
\label{eq:PQSParameter}
\xi_{||}^2 \equiv \frac{\Delta^2 F_{||}}{\FPlane}.
\end{equation}
A {\em planar squeezed state} has $\xi_{||}^2 < 1$, and also has individual component variances below the SQL, i.e.,  {$ \xi_{x}^2<  1$, and $ \xi_{z}^2< 1$ }, where $\xi_i^2 \equiv {2\Delta^2 F_{i}}/{\FPlane}$, so that $\xi_{||}^2 = (\xi_x^2 + \xi_z^2)/2$. We note that this is a metrological definition of squeezing like the Wineland criterion, in that it compares noise to the magnitude of the coherence $\FPlane$.~In order to achieve this kind of state, it is convenient to have $\langle F_{y} \rangle =0$ so that the uncertainties on the plane are only constrained by $\Delta F_{x} \Delta F_{z} \geq 0$.~The uncertainty reduction in the two planar components comes at the expense of increasing the noise in the third component, as required by $\Delta F_{z(x)} \Delta F_{y} \geq \frac{1}{2}{\langle F_{x(z)} \rangle }$. 

\section{Numerical Results}

We simulate an experiment illustrated in Figure~\ref{fig:ExperimentalScheme}.  $N_{at}=1.25\times10^6$ $^{87}$Rb atoms are cooled in the $F=1$ ground state and held in a weakly focused single beam optical dipole trap with an effective optical depth (OD) $\alpha_0 \approx 65$~\cite{KubasikPRA2009}.~To produce a PQS, we apply a magnetic field $B_y$ to coherently rotate an initially $F_x$ polarized coherent spin state in the $x,z$ plane, and stroboscopically probe the spins at four times the Larmor frequency.~We probe the atoms with $\mu s$ pulses of linearly polarised off-resonant light detuned by $\Delta =- 2 \pi \times 2$ GHz on the $D_2$ line, detected by a shot-noise limited polarimeter.~The pulses are sent through the atoms in pairs of alternating h-- and v--polarization in order to cancel the effect of tensorial light shifts~\cite{KoschorreckPRL2010b}.~Each pulse is $\tau$=1 $\mu$s long and contains $10^8$ photons, and the two pulses in a pair are separated by 1~$\mu$s.~Each stroboscopic measurement consists of a single pair of pulses with a total duration of 3~$\mu$s.

The coupling constant between the light and the atoms is $G_{1}=0.5 \times 10^{-7}$.  The atomic ensemble is initially polarized in the $x$-direction by optical pumping, i.e., into a state $\left|\ctext{+f_x}\right>^{\otimes N_{at}}$, where $\left|+\ctext{f_x}\right> \equiv \frac{1}{2} ( \left| m=-1\right> +\sqrt{2}\left| m=0\right> + \left| m=+1\right>)$ and where $N_{at}$ is subject to Poissonian fluctuations.  This last condition is not imposed by quantum physics, which would allow $N_{at}$ to be sharp, but rather represents the practical fact that trap loading is a stochastic process.  We refer to this kind of state as a~{\em Poissonian coherent spin state} (PCSS).  The $x$-polarized PCSS has average values $\langle F_{x} \rangle = \left< N_{at} \right>$,  $\langle F_{y} \rangle = \langle F_{z} \rangle =0$ and variances $\var(F_y) = \var(F_z) = \langle N_{at} \rangle/2$ (due to quantum fluctuations) and $\var(F_x) =  \langle N_{at} \rangle$ (due to $\Delta N_{at}$).  Further details of numerical simulations and the initial state are described in Ref.~\cite{ColangeloARX2013}.  The external magnetic field along $y$, $B_{y}=14.3$ mG, {$B_x=B_z =0$},  produces Larmor precession in the ($x,z$)-plane, with a  period of $T_{L} = 100$ $\mu$s, considerably longer than the $\sim \unit{1}{\micro\second}$  time required for the probing.

\begin{figure}[ht!]
\centering
\includegraphics[width=\textwidth]{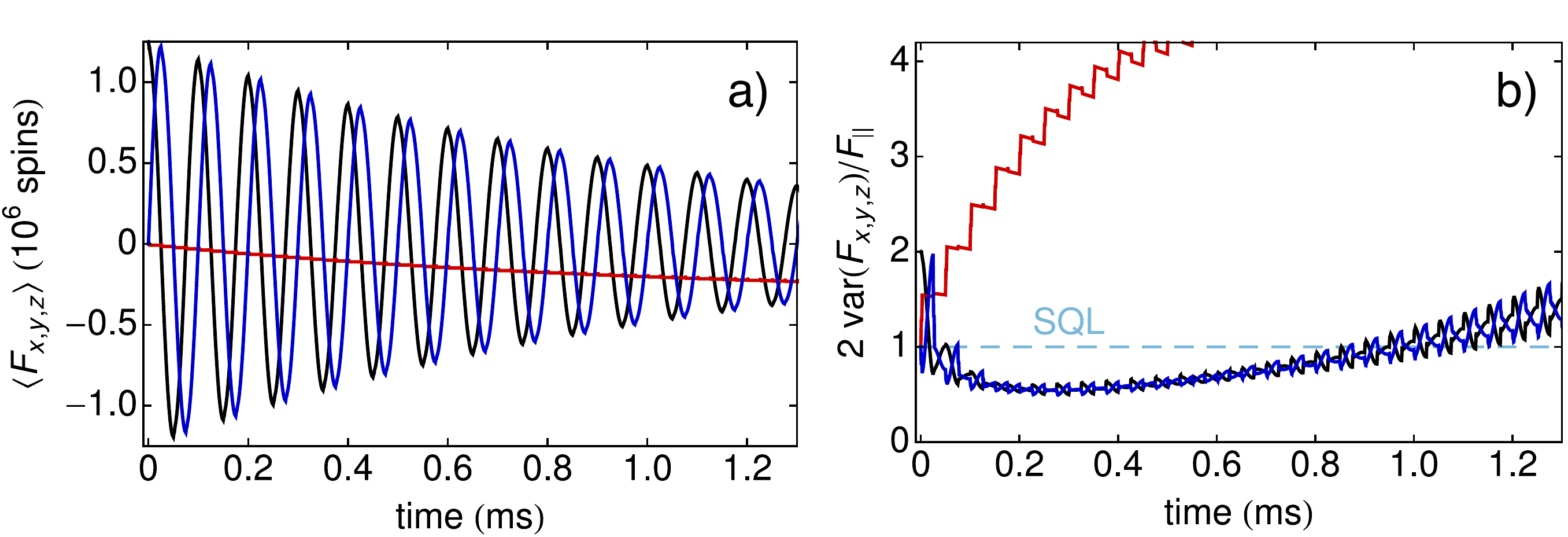}
\caption{
(a) Mean values of collective atomic angular momenta. Blue curve $\langle F_{z} \rangle$, black curve $\langle F_{x} \rangle$, and red curve $\langle F_{y} \rangle$.~The mean value in the direction orthogonal to the plane of precession remains always approximately zero $\langle F_{y} \rangle \approx 0$, as required for \pqs.~We note a reduction by roughly 40 $\%$ in the mean collective atomic spins due to optical scattering.~(b) Normalized variance of collective atomic spin components. Blue curve ${2 \Delta^2 F_{z}}/\ctext{\FPlane} $, black curve ${2 \Delta^2 F_{x}}/\ctext{\FPlane}  $, and red curve ${ 2\Delta^2 F_{y}}/\ctext{\FPlane}  $.~As expected from the spin uncertainty relations, noise on the components in the $(x,z)$-plane are drop below the standard quantum limit ${\FPlane}/2$ indicated by the dashed line.~In compensation, the out-of-plane variance $\Delta^2 F_{y}$ increases.~The discontinuities in $\Delta^2 F_{y}$ correspond to probing events. Maximal \pqs is achieved at  approximately 300$\mu$s.
The dashed blue line indicates the standard quantum limit (SQL).
\label{fig:MeanValues}}
\end{figure}

\begin{figure}[ht!]
\centering
\includegraphics[width=\textwidth]{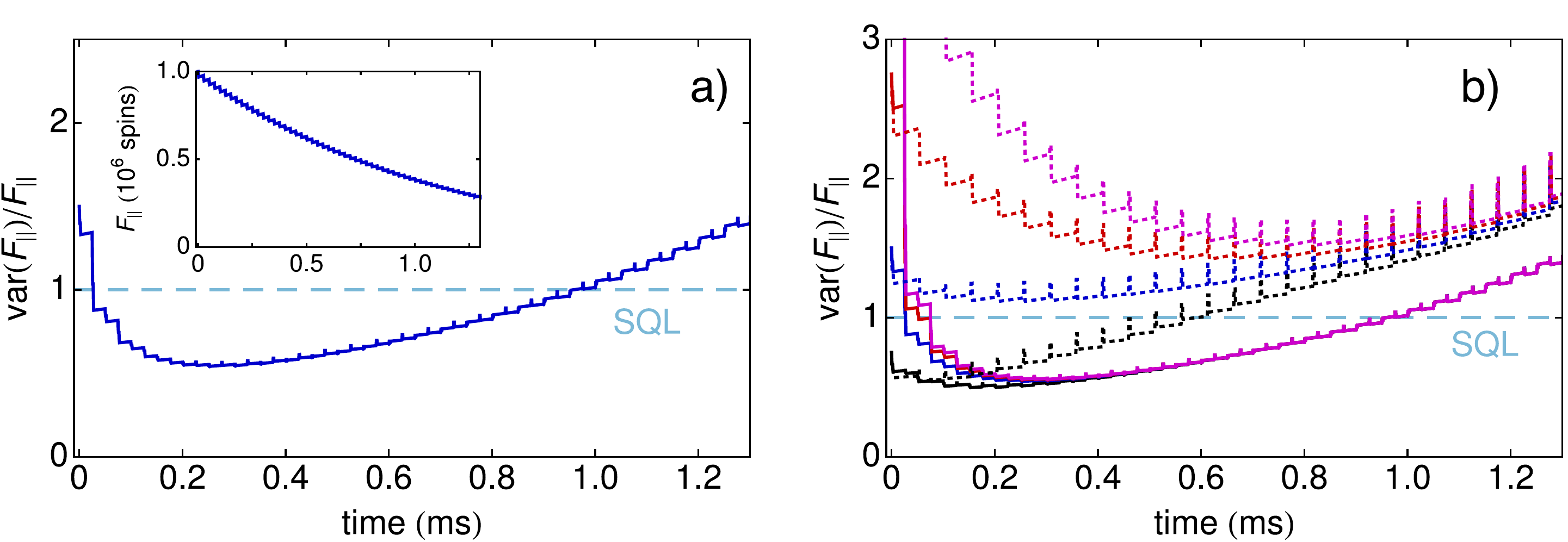}
\caption{(a) Planar squeezing parameter $\xi_{||}^2 \equiv \Delta^2 \FPlane/\FPlane$ during probing sequence. $\xi_{||}^2$ is reduced in steps due to stroboscopic probing and drops below unity, indicating planar squeezing. Maximal planar squeezing is achieved at approximately 300 $\mu$s and is reduced at longer times due to loss of atomic polarization by optical scattering. Inset shows the  {spin magnitude}  $\FPlane$, which is reduced by 40 $\%$ during the total measurement time.~(b) Planar  squeezing parameter~$\xi_{||}^2$  during probing for different initial states and different measurement strategies.~Dashed lines show stroboscopic probing on a single component ($F_{z}, -F_{z}$, or two probing events per Larmor period), and solid lines show stroboscopic measurement on two components ($F_{z}, F_{x},-F_{z},-F_{x}$ or four events).~Black, blue, red, and magenta curves correspond to different variances in atom number $\Delta^2 N_{at}= x^2 N_{at}$, $x = 1/2, 1, 3/2$ and $2$, respectively.~For the larger initial noise levels (blue, red and magenta curves) \pqs ($\xi^2_{||}<1$) can be achieved by probing both $\pm F_x$ and $\pm F_z$ (solid curves) but not by probing only $\pm F_z$ (dashed curves).~In both panels the dashed blue line indicates the standard quantum limit (SQL).
\label{fig:PlanarSqueezingParameter}}
\end{figure}

Results of numerical simulations for a typical condition are shown in Figures~\ref{fig:MeanValues} and~\ref{fig:PlanarSqueezingParameter}.~The ensemble of atoms is initially polarized in $x$-direction, and precesses in the ($x,z$)-plane due to the B-Field at the Larmor frequency.~As shown in Figure~\ref{fig:MeanValues}(a), the mean value of the out-of-plane component remains near zero, i.e., $\langle F_{y} \rangle \ll N_{at}$, so that the relevant uncertainty relation is $\Delta F_{x} \Delta F_{z} \geq |\langle F_y\rangle|/2$ imposes little back-action and the uncertainties in the ($x,z$) plane can be simultaneously reduced.~The observed deviation from $\langle F_{y} \rangle =0$ is due to residual effects of the tensorial coupling term $G_{2}(\hat{S}_{x}\hat{J}_{x}+\hat{S}_{y}\hat{J}_{y})$ in the effective Hamiltonian.

Figure~\ref{fig:MeanValues}(b) shows the variances of the spin components during the probing procedure. Because the initial state is a $F=1$ PCSS, the individual ($x,z$) variances are initially unequal, but quickly take on similar values through measurement.~They oscillate as a consequence of the Larmor precession, and are sequentially reduced due to the stroboscopic probing, which can be seen in the step-like jumps in the various curves.~At the same time, the variance in the orthogonal direction $\Delta^2 F_{y}$  increases well above the SQL, as expected for a PQS state. 

Figure~\ref{fig:PlanarSqueezingParameter}(a) shows the planar squeezing parameter $\xi_{||}^2$, which drops below unity indicating the production of \pqs by QND measurement.

\subsection{Two--spin--component measurement--induced squeezing}

The optimal measurement strategy will depend in general on the amount of initial quantum and technical noise in the angular momentum components $\Delta F_{i}(t=0)$, with $i=(x,z)$. If the initial noise in one component is sufficiently low (variance below $N_{at}$),  then it may be enough to squeeze the remaining component  by probing in  a single direction. However, in most realistic scenarios it is preferable to probe the two in-plane components, so the noise of each can be reduced.  Figure~\ref{fig:PlanarSqueezingParameter}(b) shows the development of planar squeezing for one-component and two-component measurement strategies for different initial noise conditions. 

\subsection{Achievable squeezing versus optical depth}
\label{sec:SimpleModel}
In the absence of scattering noise and decoherence, a QND measurement reduces the variance of the measured parameter by a factor $1/(1+\kappa)$, where $\kappa$ is the signal-to-noise ratio of the measurement, proportional to the number of photons used in the measurement.  At the same time, spontaneous scattering events add noise (variance) and reduce $\FPlane$, by amounts approximately linear in the number of photons used.  An often-used estimate of the trade-off between these effects~\cite{MadsenPR2004,EchanizJO2005} finds a minimum squeezing parameter $\xi^2= 1/(1+\alpha_0 \eta) + 2\eta$, where $\eta$ is the probability for any given atom to suffer a spontaneous emission event.   This expression has a minimum $\xitwomin=2\sqrt{(2/\alpha_0)}$ for the {optimal} $\eta$, which is  $\eta_{0}=1/\sqrt{2 \alpha_0}$. 
For a typical system with optical density $\alpha_0 \approx 25$ this optimal value gives a lower bound on the amount of squeezing given by roughly $\xitwomin \approx 0.5$ (3dB of noise reduction)~\cite{HammererPRA2004}. 

\begin{figure}[ht!]
\centering
\includegraphics[width=\textwidth]{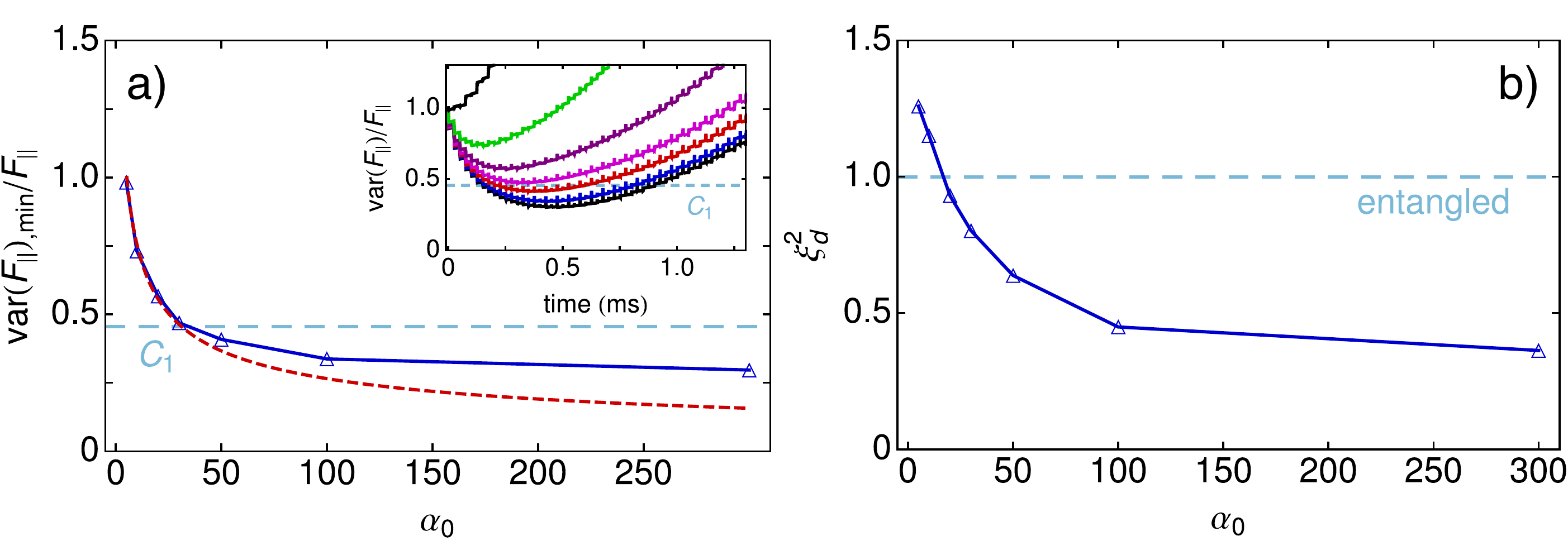}
\caption{Spin squeezing and entanglement of PQS.
(a) Entanglement detection in planar squeezing using the separability criterion $\Delta^2 \FPlane/{N_{at}} \ge C_F$ of He, et al.~\cite{HePRA2011},  for which $C_1 = 7/16$.  Inset shows $\Delta^2 \FPlane/{N_{at}}$ versus time for  $\alpha_0=5$ (black), 10 (blue), 20 (red), 30 (magenta), 50 (purple) 100 (green), and 300 (black).~Main graph shows minimal values achieved versus $\alpha_0$ (blue symbols).  Blue line is a guide to the eye.~The criterion detects entanglement for $\alpha_0 \gtrsim 35$.~Dashed red line: Simple estimate of planar squeezing parameter $\xi_{||,\mathrm{min}}^{2}$, for optimal spontaneous emission probability $\eta_0=1/\sqrt{2 \alpha_0}$, see Section \ref{sec:SimpleModel}.~(b) The spin squeezing parameter of Eq.5(d) of Ref.~\cite{VitaglianoPRL2011} detects entanglement for $\xi_d^2<1$ (see Eq.~(\ref{eq:VitaglianoSSIneq3})).~Again, we plot the minimal values achieved versus $\alpha_0$ for the curves in the inset of (a). \label{fig:EntanglementDetection}
}
\end{figure}

\gtext{The predicted planar squeezing by this simple model is shown in Figure~\ref{fig:EntanglementDetection}(a) (red dashed line), for an optical depth ranging from  $ 0 < \alpha \le 300$.  Also shown in  Figure~\ref{fig:EntanglementDetection}(a) are full simulation results (inset), from which we can extract a more accurate estimate of the achievable planar squeezing, shown as blue symbols in the main graph.  The difference at large $\alpha_0$ (and thus large squeezing) may be attributable to finite precession angle between the h- and v-polarized parts of the probing.  In the next section, we characterize the entanglement content of the generated planar squeezed states.}

\section{Spin squeezing inequalities and entanglement criteria}

A number of spin-squeezing inequalities can in principle detect entanglement in planar squeezed states.  
He et al.~\cite{HePRA2011} give a simple inequality, obeyed by all separable states:
\begin{equation}
\label{eq:HeSSIneq}
\frac{\Delta^2 \FPlane}{N_{at}} \ge C_{F},
\end{equation}
where $C_1 = 7/16$.~As shown in Figure~\ref{fig:EntanglementDetection}(a), our procedure for producing planar squeezing can violate this inequality for optical depths $\alpha_0\gtrsim 35$. 

A generalized spin squeezing parameter from  Vitagliano et al.~\cite{VitaglianoPRL2011} (Eq. 5(d))
\begin{eqnarray}
\xi_d^2\equiv \frac{(N_{at}-1)[(\tilde{\Delta}F_x)^2+(\tilde{\Delta}F_z)^2]}{\left<\tilde{F}_y^2\right>-N_{at}(N_{at}-1)F^2} \,
\label{eq:VitaglianoSSIneq3}
\end{eqnarray}
where $\left<\tilde{F}_i^2\right>\equiv\left<F_i^2\right>-\left<\sum_n (f_i^{(n)})^2\right>$ and  $(\tilde{\Delta}F_i)^2=(\Delta F_i)^2-\left<\sum_n (f_i^{(n)})^2\right>$ are modified second order moments and variances, also detects entanglement ($\xi_d^2<1$) in planar squeezed states, as shown in Figure~\ref{fig:EntanglementDetection}(b).

Regarding the difficulty of detecting entanglement in PQS states, we note a significant difference relative to one-component spin squeezing.~With just one component, the border between squeezed and un-squeezed states coincides with the border between separable and entangled states.~That is, a CSS  is a pure product state and also has SQL noise.~In contrast, by the planar squeezing criterion of He et al.~\cite{HeNJP2012} (given as Eq.~(\ref{eq:PQSParameter}) ) a CSS (with a definite $N_{at}$), is already a planar squeezed state with $\xi_{||}<1$.~In this sense, planar squeezing is easier to achieve than entanglement.  

\section{Applications: Optical magnetometry}

We now describe a possible application of PQS in optical magnetometry,~for determination of arbitrary angles with precision below the SQL.~States that are squeezed in only one component can give a metrological advantage over a limited range of angles.~Protocols to employ these states either require prior knowledge of the phase (enough to place them within the range of improved sensitivity), or adaptive procedures to determine the phase during the measurement.~In contrast, planar squeezed states can give improved sensitivity for any precession angle~\cite{HePRA2011,HeNJP2012}.~Here we show that QND-generated PQS states are effective for this purpose.

\begin{figure}[!ht]
\centering
\includegraphics[width=\textwidth]{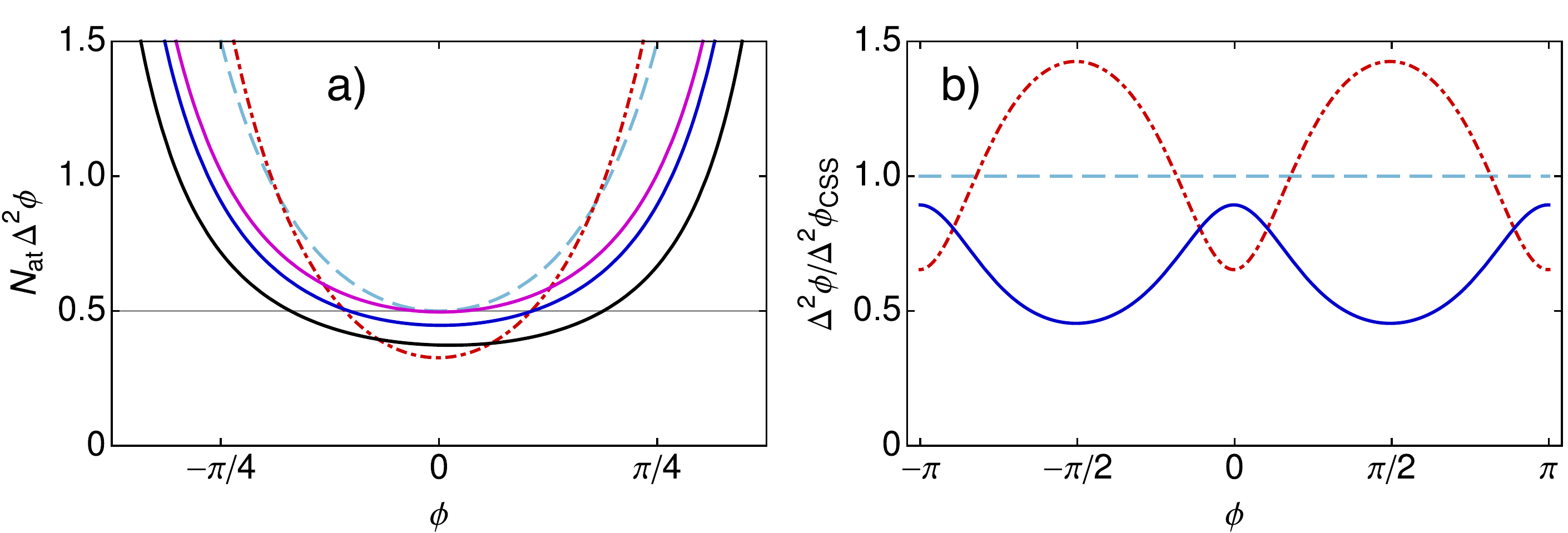}
\caption{
(a) Phase-estimation variance $\Delta^2 \phi$ as a function of precession phase $\phi$ for planar squeezed states generated as described above.  SQL $\Delta^2 \phi = 1/(2 N_{at}) $ is indicated by the grey horizontal line, while curves show phase estimation uncertainty for various states.  Black, blue, and magenta solid curves show planar squeezed states generated as in Figure~\ref{fig:PlanarSqueezingParameter}(b) from coherent spin states with pre--squeezing number uncertainties $\Delta^2 N_{at}= x^2 N_{at}$, $x = 1/2, 1$ and $2$, respectively.
For such realistic PQS states, it is possible to reconstruct an arbitrary parameter with precision below the SQL within a significant interval of $\phi$.
The blue dashed curve shows the PCSS state.
The red dot-dashed curve shows a single-component squeezed state (SSS) produced by probing only the $F_z$ component of the atomic spin (blue dashed curve of Figure~\ref{fig:PlanarSqueezingParameter}(b)).
This state has a $\sim50\%$ reduction of the variance of the $F_z$ component of the atomic spin, and a $\sim20\%$ reduction of the collective spin coherence.
It gives improved precision in a smaller region around the phase $\phi=0$.
The asymmetry of the PQS and SSS curves is a result of the non-zero cov$(F_x,F_z)$ due to imperfect cancellation of tensorial light shifts during the measurement. 
(b) Comparison of phase estimation variance $\Delta^2 \phi$ for a planar squeezed state (solid blue line),  a single component squeezed state (dot-dashed red line), and the PCSS (light blue dashed line).~States were generated by QND measurement as above, but with $\alpha_0 = 60$. All curves are normalized to the PCSS value.~A single component squeezed state can beat the precision of the PQS state for a few particular phases, but PQS states are more precise on average and offer an advantage relative to the PCSS for any phase.
\label{fig:DeltaPhi}}
\end{figure}

We consider a state, initially oriented in the $F_x$ direction, and allowed to precess in response to a field ${\bf B}$ along the $y$-direction.  The precession angle versus time $t$ will be ${\phi}=\omega_{L} t$, where $\omega_{L}$ is the Larmor frequency, which in turn is proportional to  $B_{y}$. 
After precession, the measured component $F_z$ is
\begin{equation}
F_z\supout = F_x\supin \sin \phi + F_z\supin \cos \phi,
\end{equation}
where superscritps $\supin, \supout$ indicate the operators before and after the precession, respectively.
We can calculate the uncertainty in $\phi$ as $\Delta^2 \phi = { {\Delta^2 F_z\supout}}/{|d \left< \right. F_z\supout \left.\right>/d \phi|^2}$ or, given that $\left<\right.{F_z\supin}\left.\right> = 0$, 
\begin{eqnarray}
\label{eq:DelPhi}
\Delta^2 \phi& = & \frac{{\Delta^2 F(\phi)}} { |F|^2 \cos^2 \phi }
\end{eqnarray}
where $\Delta^2 F(\phi)\equiv \Delta^2 F_{x}\supin\sin^2\phi +\Delta^2 F_{z}\supin\cos^2\phi + \mathrm{cov}(F_{x}\supin,F_{z}\supin)\sin2\phi$, and $ \cov(A,B) \equiv  \frac{1}{2} \langle A B + BA  \rangle - \langle A \rangle \langle B \rangle$ is the covariance.~We note that PQS states reduce the planar variance for arbitrary angles on a finite interval, with the exception of the specific singular values which make the denominator in Eq.~(\ref{eq:DelPhi}) equal to zero. This feature is depicted in Figure~\ref{fig:DeltaPhi}(a), for realistic PQS states with different levels of technical noise, generated via QND measurements. 

Figure~\ref{fig:DeltaPhi}(b) shows a comparison of the phase variances, relative to the PCSS. 
The results show that while a single-component squeezed state is more precise than the PCSS (and also the PQS) over a range of angles, the PQS gives a near-uniform advantage relative to the CSS and can be used for phase reconstruction below the SQL without prior knowledge of $\phi$. 

\section{Conclusions}

We have studied numerically the possibility to generate the recently-proposed ``planar quantum squeezing,'' in which the variances of two orthogonal spin components are simultaneously squeezed, via quantum non-demolition measurement of cold atomic ensembles.~We find that significant planar squeezing can be generated under realistic conditions and that this squeezing implies entanglement detectable with spin-squeezing inequalities.~Considering the use of planar squeezed states in an optical magnetometry context, we find that planar squeezing can give a metrological advantage for estimation of arbitrary phase angles, whereas single-component spin squeezing is only advantageous for specific angle ranges.~This is promising for high-bandwidth atomic magnetometry, in which a changing precession frequency may make it difficult or impossible to anticipate the precession phase.

\bigskip

\section*{ACKNOWLEDGEMENTS}

The authors gratefully acknowledge G\'{e}za T\'{o}th for fruitful discussions. They also would like to thank Naeimeh Behbood, Ferran Martin Ciurana and Mario Napolitano for helpful discussions. This work was supported by the Spanish MINECO under the project MAGO (ref. no. FIS2011-23520), by the European Research Council under the project AQUMET and by Fundaci\'{o} Privada Cellex.  G.P. acknowledges financial support from Marie-Curie International Fellowship COFUND. 

\bigskip

\section*{References}

\bibliographystyle{iopart-num}
\bibliography{PQSreferences}

\end{document}